\documentclass[preprint,12pt,3p,review]{elsarticle}

\usepackage{amssymb}
\usepackage{graphicx,amsmath}
\usepackage{natbib,lineno}
\biboptions{numbers,sort&compress}

\usepackage[breaklinks,colorlinks=true,citecolor={blue}, linkcolor= {blue}, urlcolor ={blue}]{hyperref}
\journal{Elsevier}

\begin{document}
\begin{frontmatter}
\title{Optical absorption of a Cu$_2$SnS$_3$ (CTS) layer trapped by metallic thin films in multilayer configuration}


\author[label3,label4]{Le Tri Dat}
\address[label3]{Engineering Research Group, Dong Nai Technology University, Bien Hoa City, Vietnam}
\address[label4]{Faculty of Engineering, Dong Nai Technology University, Bien Hoa City, Vietnam.}
\ead{letridat@dntu.edu.vn}

\author[label8,label9]{Nguyen Duy Vy\corref{cor1}}
\address[label8]{Laboratory of Applied Physics, Science and Technology Advanced Institute, Van Lang University, Ho Chi Minh City, Vietnam}
\address[label9]{Faculty of Applied Technology, School of Technology, Van Lang University, Ho Chi Minh City, Vietnam}
\ead{nguyenduyvy@vlu.edu.vn} \cortext[cor1]{Corresponding author.}


\begin{abstract}
Coating and reflecting thin films for energy harvesting purposes are interesting topics in both theoretical and experimental research. The thin film could help to enhance the absorption of the system via its specific optical properties depending on the optical wavelength and the stacked layer thickness. Here, by using Maxwell's equations for the electromagnetic fields penetrating thin films, we examined in detail the absorption of a CTS layer coated by nanometer-thick thin films of several materials, Au, Ag, Cu, Al, and figured out the optimal thickness range for the outer layers of the solar cell to optimize thermal energy harvesting from the light. In particular, the absorption has been shown to be significantly enhanced thanks to the optical cavity effect, and the maximal absorption of the system could reach 60\% for ... These results could help in suitably choosing the detailed thickness for the structure of the solar cell and other energy harvesting objects.
\end{abstract}

\begin{keyword}
metallic absorption \sep solar cell \sep optical cavity \sep Maxwell's equation \sep film thickness
\end{keyword}

\end{frontmatter}

\section{Introduction}
Optical energy harvesting from solar radiation remains a critical research focus, driving decades of innovation in photonic materials and device architectures for enhanced light absorption across diverse applications \cite{KatLaser16,PahujaAPYA20,LiMaRev21,LiuPowderRev21, GirtanOptMat22}. While sophisticated approaches utilizing metallic thin films \cite{DotanNmat12,KatsAPL14,ParkOL15}, lossy films on reflectors \cite{KatsNews14}, plasmonic nanoparticles \cite{LeeOE10,SpinelliJOpt12, SrivastavaAPYA18Au, HamedAPYA20Ag}, graphene \cite{JabeenIEEEQm18}, and nanogratings \cite{SubhanRSC20} have significantly advanced absorption control from THz to visible wavelengths, the quest for efficient, stable, and sustainable absorber materials persists. 
The fundamental mechanism bases on converting absorbed photons into useful energy in the form of thermal or electrical currents. Opto-electrical materials achieve this by exciting charge carriers directly \cite{BucherAPYA78}, making the absorber's efficiency important, as it determines the fraction of incident solar energy harnessed by the device. Consequently, systems combining high absorption ratios with exceptional environmental stability are highly sought after.

\begin{figure}[!h] \centering
\includegraphics[width=0.64\textwidth]{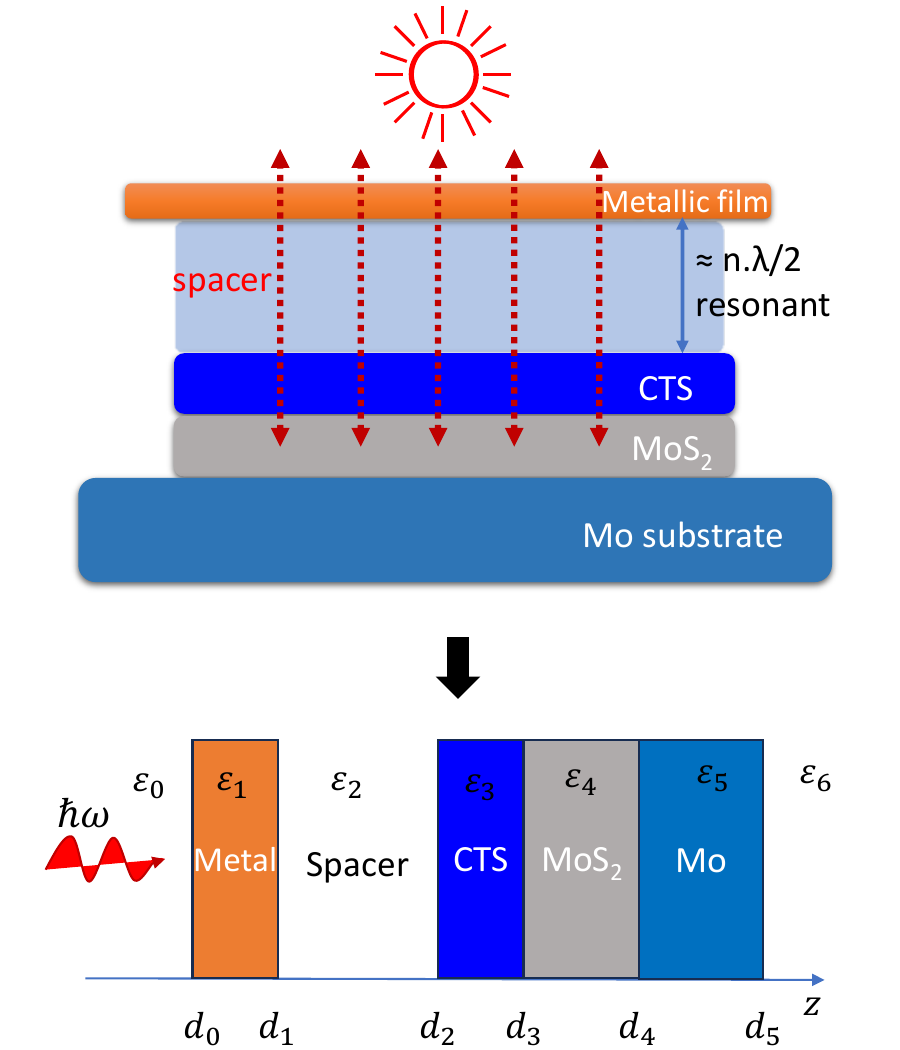}
\caption{Model of a CTS layer embeded inside a multi-layer system parallelly arranged. The electromagnetic components of the $i$th layer are presented as $E^+_i$ and $E^-_i$ for the incoming and outgoing fields, respectively. The spacer thickness is approximately equal to a number of half-optical wavelength, $d_2-d_1\simeq n\lambda/2$.} \label{fig1model}
    \end{figure}
Emerging as a highly promising photovoltaic absorber material, Copper Tin Sulfide (CTS, Cu$_2$SnS$_3$) offers an effective combination of properties: an ideal near-infrared direct bandgap ($\sim$1.0--1.5 eV), an extremely high absorption coefficient ($>10^4$ cm$^{-1}$), and constituent elements that are earth-abundant and low-toxicity \cite{Birkmire1997rev, Lin2023energy, Lu2024Nanoscale}. These attributes make CTS particularly suitable for thin-film solar cells, where efficient light capture in ultrathin layers is essential. Recent studies highlight its potential for defect tolerance and tunable optoelectronic properties, contributing to steadily increasing power conversion efficiencies \cite{Ghosh2023}. Integrating CTS into resonant cavity structures presents a promising avenue to further exploit its intrinsic absorption strength through optical field enhancement. Previous studies \cite{Vy2024AIP} have shown that by using a reflective layer, the absorption of the cavity system could be greatly enhanced. Therefore, it creates a motivation for using CTS as an element of the cavity system and it is expected that the optical absorption could be highly obtained.

In this study, we investigate the integration of CTS into a cavity-enhanced absorption system, as depicted in Fig. \ref{fig1model}. CTS is positioned to function either as a primary absorber layer, potentially replacing metallic films, or as a complementary layer supporting metallic components. We employ analytical solutions derived from Maxwell's equations to model the electromagnetic field distribution and optical absorption within the multilayer structure, explicitly including CTS layers. The system's geometry, particularly the cavity spacing relative to the wavelength, is optimized to induce resonant field enhancement between the reflective surface (metallic or other) and the CTS layer, thereby maximizing absorption within the CTS. This approach aims to leverage both the superior material properties of CTS and the intensity amplification offered by cavity resonance. The resulting design principles provide concrete guidance for fabricating high-performance, CTS-integrated solar energy harvesting devices.

\section{Theoretical model and background}
Exerted by an electromagnetic field of energy $\hbar\omega$, the first metallic layer partly reflects the incoming electric field. The field inside a layer is presented as follows
\begin{align}
E = E^+e^{ikz} + E^-e^{-ikz}, \label{elec_field}
\end{align}
where, $E^+$/$E^-$ is the incident/reflected field, and $k$ is the wave number. The electric and magnetic fields are written as follows,
\begin{align}
E &= E^+e^{ikz} + E^-e^{-ikz},\\
H &= H^+e^{ikz} + H^-e^{-ikz},    \label{eq_fields}
\end{align}
where $E^+$ and $H^+$ are the incident fields, $E^-$ and $H^-$ present the reflected fields, and $k = \sqrt{\epsilon} \omega/c$ is the wave number. Considering a position $d_j$ that was a joint between two layers, we obtained a pair of equations at that point.
\begin{align}
E_j^+e^{ik_jd_j} + E_j^-e^{-ik_jd_j} &= E_{j+1}^+e^{ik_{j+1}d_j} + E_{j+1}^-e^{-ik_{j+1}d_j} \\
H_j^+e^{ik_jd_j} + H_j^-e^{-ik_jd_j} &= H_{j+1}^+e^{ik_{j+1}d_j} + H_{j+1}^-e^{-ik_{j+1}d_j},\label{eq_interface}
\end{align}

In this study, we assumed $\Vec{k} \parallel \Vec{z}$, so $|\Vec{k}| = k_z = k$. One has $2(N-1)$ equations for $N$ different layers. However, to calculate conveniently, we converted them to a matrix equation with the size $M\times M$ where $M$ equals $2(N-1)$ as used in Ref. \cite{DatOptCom22}. The matrix has the form,    
\begin{equation}\label{mat_1}
\left[\begin{matrix}
e^{-ik_0 d_0} & -e^{ik_1 d_0} & -e^{-ik_1 d_0} & ... & 0 \\
-e^{-ik_0d_0} & -\frac{k_1}{k_0}e^{ik_1d_0} & \frac{k_1}{k_0}e^{-ik_1 d_0} & ... & 0 \\
... & ... & ... & ... &...\\
0 & ... & e^{ik_{N-1}d_{N-1}} & e^{-ik_{N-1}d_{N-1}} & -e^{ik_{N}d_{N-1}} \\
0 & ... & e^{ik_{N-1} d_{N-1}} & -e^{-ik_{N-1}d_{N-1}} & -\frac{k_N}{k_{N-1}}e^{ik_{N}d_{N-1}} \\
\end{matrix}
   \right]
   \left[
    \begin{matrix}
            E^-_0\\
            E^+_1\\
            E^-_1\\
            .\\
            .\\
            .\\
            E^+_{N}
    \end{matrix}\right] = \left[
    \begin{matrix}
            -1 \\
            -1 \\
            0 \\
            .\\
            .\\
            .\\
            0
    \end{matrix}\right],
\end{equation}
where, it was assumed $E^+_0 = 1$ and $E^-_N = 0$. One can solve Eq. (\ref{mat_1}) and obtain these fields for each layer.


\subsection{Absorption of a metallic layer}  
The MoS$_2$ plays the role of a reflective layer located behind the thin film and gives rise to an enhancement of fields by partly or totally reflecting the light to the CTS film. The absorption of a certain layer of thickness $d$ is calculated as follows
\begin{align}
P_{abs} = \int_{d_i}^{d_{i+1}} dz\left| E_i(z)\right|^2 \textbf{Im}[1 - \epsilon(\omega)]\frac{\omega}{c}, \label{eq_metal_absorp}
\end{align}
where $E_i(z)$ is the electric field inside the film $i$th, $\epsilon(\omega) $ is the complex optical dielectric function with optical frequency, $\omega = 2\pi c/\lambda$, and the integration is taken over the film thickness $[d_{i}, d_{i+1}]$ as shown in Fig. \ref{fig1model}(bottom). We use the Lorentz-Drude model \cite{VialPRBAu05,VialAPYA11} for the dielectric function of the metallic layer, 
%
\begin{align}
\epsilon_{LD}(\omega) = \epsilon_b(\omega) + \epsilon_f(\omega),\label{dielectric}
\end{align}
where $\displaystyle\epsilon_f(\omega) = 1 - \frac{f_0\omega_p^2}{\omega^2 + i\Gamma_0\omega}$ is the intraband part and $\displaystyle \epsilon_b(\omega) = \sum_{j=1}^K\frac{f_j\omega_p^2}{(\omega_j^2-\omega^2) + i\Gamma_j\omega}$ describes the interband part of the dielectric function, which could be written as $\epsilon(\omega) = \epsilon_{LD}^1-i\epsilon_{LD}^2$. 

\subsection{Absorption of a metallic layer with a reflective layer.}    
We assume a configuration that the reflective layer lies below the CTS layer Fig. \ref{fig1model}(top) and has the thickness of $d_{43}=d_4-d_3$. Together with the CTS layer, this reflective layer creates an optical cavity of length $L_C =\delta=n\lambda/2$ and could enhance the electric field stored inside the cavity up to several factors \cite{VyAPL13,VyAPEX15, VyAPEX16, VinhPham2020apex}. To figure out the electric fields, a matrix equation was written out as follows
\begin{equation}\label{eq_mat_2}
\left[\begin{matrix}
1 & -1 & -1 & 0 \\
-1 & -\frac{k_1}{k_0} & \frac{k_1}{k_0} & 0 \\
0 & e^{ik_1d} & e^{-ik_1d} & -e^{ik_2d}\left(1 + e^{i2k_2\delta} \right) \\
0 & e^{ik_1d} & -e^{-ik_1d} & -\frac{k_2}{k_1}e^{ik_2d}\left(1 - e^{i2k_2\delta} \right) \\
        \end{matrix}
        \right]\left[
\begin{matrix}
            E^-_0\\
            E^+_1\\
            E^-_1\\
            E^+_2
\end{matrix}\right] = \left[
\begin{matrix}
            -1 \\
            -1 \\
            0 \\
            0
\end{matrix}\right].
    \end{equation}
Here, at the position $z = d + \delta$, the field is assumed ${E_{2}^- = E_{2}^+e^{i2k(d+\delta)}}$ with $L_C = \delta$ to obtain the total reflection. 
   
\section{Results and discussions}
In constructing our multilayer solar cell system, precise modeling of optical properties is critical for performance optimization. For the metallic layers (e.g., Au, Ag), which serve as reflective contacts and plasmonic enhancers, we adopt the Lorentz-Drude model parameters from Rakic et al. Ref. \cite{RakicAO98}). This widely validated dataset accounts for frequency-dependent dispersion and interband transitions, essential for simulating noble metals' behavior across visible to near-infrared wavelengths. The CTS absorber layer exhibits complex dielectric responses critical for photon harvesting. Among numerous theoretical and experimental studies, we utilize the dielectric function from Crovetto et al. \cite{Crovetto2016}, which used first-principles calculations with spectroscopic ellipsometry measurements, ensuring accuracy for thin-film configurations. The spacer layer is idealized as air ($\epsilon_{\text{air}} = 1$), a simplification that neglects dispersion but validly represents void-based architectures or inert-gas environments in experimental setups. For the MoS$_2$ dielectric interlayer—a 2D transition metal dichalcogenide offering excitonic enhancement and interface passivation—we employ the anisotropic complex refractive index from Song et al. \cite{Song2018}, derived from monolayer-specific spectroscopy to capture its in-plane/out-of-plane optical anisotropy. Finally, the molybdenum (Mo) back-contact layer, chosen for its stability and ohmic properties, adopts experimental dielectric constants from Kirillova et al. \cite{Kirillova1971}, which provide temperature-invariant bulk values relevant to thick-film deposition techniques.

\subsection{Electric field distribution for various metallic coatings}
\begin{figure}[!h] 	\centering
\includegraphics[width=.7\textwidth]{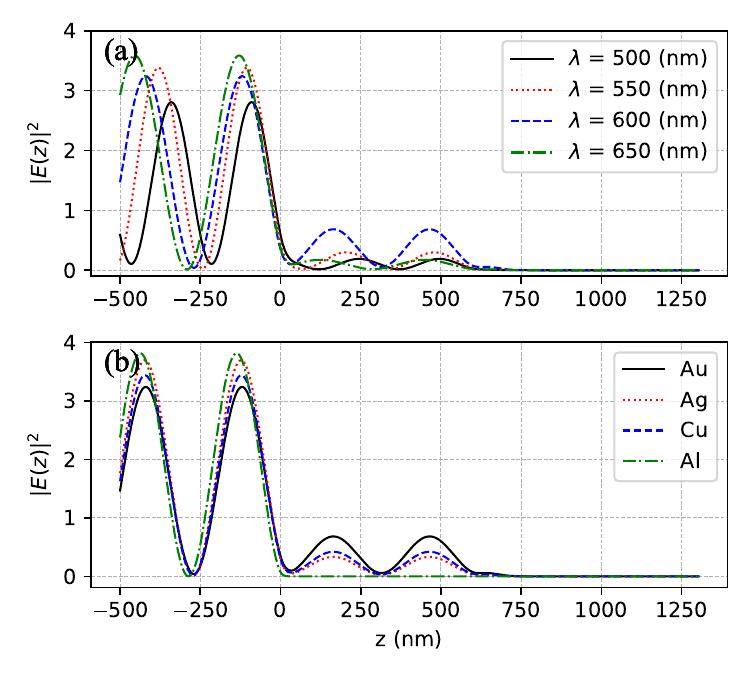}
\caption{The field distribution in various regions of the system normalized to the incoming field. The first metallic thin film (green) together with the second (violet) gives rise to an optical microcavity and the constructive/destructive pattern inside the cavity. The wavelength of the input laser is $\lambda$ = 633 nm and the cavity length is $\delta=\lambda$.} \label{fig_E2_metals}
\end{figure}
First, the electric field distribution for Au coating is shown in Fig. \ref{fig_E2_metals}(a) with the following parameters: Au film thickness $d_1-d_0$ = 47 nm, spacer thickness $d_2-d_1$ = $n\lambda/2$ + 565 nm, CTS layer thickness 100 nm, and bulk Mo substrate thickness 500 nm. The results show that the incoming field at $\lambda$ = 600 nm (blue dashed line) exhibits the greatest enhancement compared to other fields. This enhancement arises from system resonance, where the effective cavity length $L_c=d_2-d_1$ matches the optical wavelength $\lambda$ according to $L_c+\Delta\simeq n\lambda$, with $\Delta\sim$ 35 nm representing the field penetration depth into the metallic layer.

To investigate the influence of coating materials on field amplification, Fig. \ref{fig_E2_metals}(b) displays $|E|^2$ inside and outside the system. Here, the laser wavelength is $\lambda$ (He-Ne laser) = 633 nm, the cavity length is $L_C=\lambda$, and the metallic layer thickness is 30 nm. Au demonstrates the strongest field enhancement among all tested metals.


\subsection{Optical absorption of CTS solar cells}
\begin{figure}[!h] 	\centering
\includegraphics[width=0.7\textwidth]{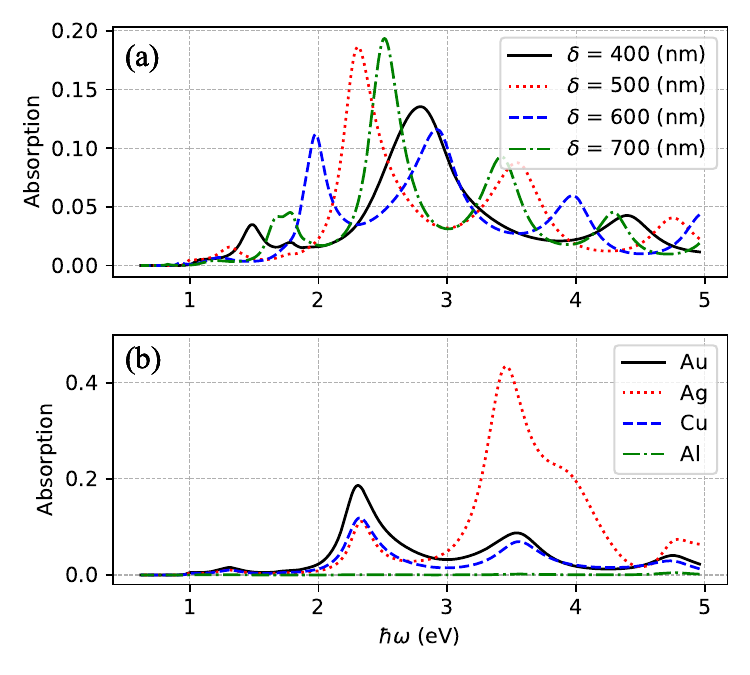}
\caption{(a) Simulated absorption spectra of Cu$_2$SnS$_3$ (CTS) thin films as a function of cavity length (spacer thickness), showing resonance-enhanced absorption for thicknesses of 400 nm, 500 nm, 600 nm, and 700 nm. The absorption peaks shift and intensify depending on the optical path length, with optimal enhancement occurring near the CTS bandgap.
(b) Absorption spectra of CTS with different back reflector metals (Au, Ag, Cu, Al) at a fixed, optimized cavity thickness. Noble metals (Au and Ag) exhibit stronger absorption enhancement due to higher reflectivity and plasmonic effects, improving light confinement in the active layer.} \label{e2_CTS}
\end{figure}

Figure \ref{e2_CTS} presents the simulated optical absorption characteristics of Cu$_2$SnS$_3$ (CTS) thin films in a metal-spacer-semiconductor cavity configuration. Panel (a) illustrates the variation in absorption spectra as a function of spacer (cavity) thickness, where four distinct thicknesses—400 nm, 500 nm, 600 nm, and 700 nm—are considered.
The metallic coating is assumed to have the thickness of 47 nm \cite{OptCommun17}, that of CTS is 92 nm, of MoS$_2$ is 100 nm. The optical wavelength is from 250--2000 nm.
The absorption intensity exhibits a clear resonance behavior, with peak positions and magnitudes sensitive to the cavity length. Notably, the optimal absorption enhancement occurs around photon energies of 1.2 to 1.4 eV, which closely aligns with the direct bandgap range of CTS reported in the literature. This finding is consistent with earlier reports by He et al. \cite{He2017} who reported  0.97 eV range for CTS depending on the crystalline phase and preparation conditions,
by Hossain et al. \cite{Hossain2022} of 0.9 eV in high-quality CTS thin films after optimized sulfurization,
and by Laghchim et al. \cite{Laghchim2023} of 1.1--1.18 eV as optimal for photovoltaic performance, matching well with our current design. 

Figure \ref{e2_CTS}(b) explores the role of various back-reflector metals--Au, Ag, Cu, and Al--on the absorption efficiency of CTS at a fixed, optimized cavity thickness. It is observed that gold (Au) leads to the most pronounced enhancement in absorption, followed by silver (Ag), copper (Cu), and aluminum (Al). This trend can be attributed to the superior plasmonic properties and higher reflectivity of Au and Ag in the near-infrared and visible regions. The presence of such metals enhances the optical field intensity inside the cavity via constructive interference and plasmonic resonance, thereby boosting light trapping and absorption in the CTS layer. This observation is consistent with the photonic cavity designs adopted in theoretical works such as those by Mathew \cite{Mathew2025} and Laghchim \cite{Laghchim2023}, where interface engineering and dielectric-metal coupling were found to significantly influence light absorption and device efficiency.

The cavity-induced enhancement is also analogous to the thickness-dependent absorption behavior observed in experimental studies. For instance, Suryawanshi et al. \cite{Suryawanshi2021} reported that increased CTS thickness in chemically deposited films resulted in improved crystallinity and reduced bandgap, thereby enhancing photoelectrochemical response. Similarly, numerical investigations by Kutwade et al. \cite{Kutwade2021} and Rahaman et al. \cite{Rahaman2025} indicated that absorption and carrier generation are maximized in CTS devices when the absorber thickness is in the range of 500--700 nm, further supporting the observed cavity effect.

Therefore, Fig. \ref{e2_CTS} shows that the optical absorption in CTS thin films can be greatly enhanced by optimizing the cavity length and stacking noble-metal reflectors. These enhancements are critical for improving the performance of CTS-based photovoltaic and optoelectronic devices, and align well with the design strategies discussed in recent works. The tunability of the cavity absorption peak near the intrinsic bandgap of CTS implies the potential of cavity engineering in overcoming the relatively modest absorption efficiency typically seen in undoped CTS thin films.

\begin{figure}[!ht] 	\centering
\includegraphics[width=0.7\textwidth]{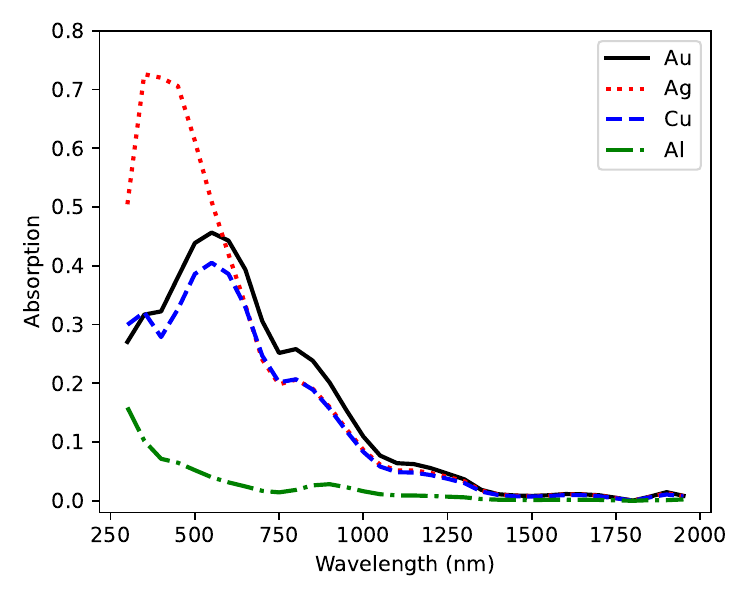} 
\caption{The absorption of the CTS structure was calculated under various configurations. The maximum absorption values for CTS with four different metallic layers (Au, Ag, Cu, and Al) are presented. Specifically, for Ag, the absorption reaches nearly 75\% at a wavelength of 350 nm when the metal thickness is $t_m = 20$ nm and the spacer thickness is $\delta = 350$ nm. In comparison, the absorption peaks at 45\% for Au with $\delta = 230$ nm, and at 40\% for Cu with $\delta = 310$ nm, both at a wavelength of 550 nm and the metallic thickness $t_m = 20$ nm. } \label{absorption_cts}
\end{figure}

In Fig. \ref{absorption_cts}, the maximal absorption has been presented where we scan a wide range of optical energy to find the wavelength that the absorption is maximized. It has been shown that the maximal absorption is from Ag coating (red dotted line) over a wide spectral range (250–2000 nm), followed by Au coating (black solid line), and Al coating (green dash-dotted line) gives the lowest values.
The architecture effectively forms an asymmetric Fabry–Pérot-like cavity, where interference, multiple reflections, and plasmonic effects enhance light–matter interactions in the CTS layer.

This is due to Ag's high reflectivity and low optical loss, particularly in the visible and near-infrared (NIR) regimes. Its complex dielectric function exhibits a low imaginary part and a plasma frequency well-positioned for visible-light enhancement, facilitating strong surface plasmon resonance (SPR) effects that concentrate the electromagnetic field inside the cavity and elevate the photon absorption probability in the adjacent CTS layer \cite{Maier2007, Atwater2010}.

Gold (Au), while often used in plasmonic systems for its chemical stability and excellent conductivity, exhibits strong interband transitions near 500–600 nm that introduce additional absorption losses and reduce the effectiveness of SPR coupling in the visible range \cite{JohnsonOptConstPRB}. These losses make Au slightly less effective than Ag in broadband absorption enhancement, a trend also observed in other photonic and photovoltaic cavity systems \cite{Atwater2010, Tan2012}. Meanwhile, copper (Cu)—an earth-abundant and low-cost plasmonic metal—shows moderate absorption enhancement but suffers from higher intrinsic damping due to higher imaginary dielectric constants, as shown in Rakic’s optical data \cite{RakicAO98}. Its performance is further affected by its tendency to oxidize and form rough interfaces, both of which degrade plasmonic quality \cite{Brongersma2015}.

Aluminum (Al), although attractive for its low cost and high reflectivity in the ultraviolet (UV), supports SPR predominantly in the UV and short visible range. Furthermore, Al exhibits substantial ohmic losses and forms a native oxide layer that suppresses plasmonic effects in the visible-NIR range, leading to the weakest absorption enhancement among the metals studied \cite{Knight2013}.

The inclusion of a MoS$_2$ interlayer beneath the CTS absorber introduces additional benefits. MoS$_2$ is known to provide favorable band alignment for hole transport and can act as a buffer or charge transfer layer in heterostructured solar cells \cite{Zou2023}. Its refractive index and weak excitonic resonances in the visible regime may contribute to improved optical impedance matching, thereby reducing reflection and enhancing absorption in the active layer. Recent studies have demonstrated the effectiveness of Mo$_2$ in improving charge separation and extraction in similar layered structures, making it a promising partner for CTS \cite{Tan2017}.

In broader context, these results underscore the importance of metal selection in photonic and plasmonic absorber designs. Ag has consistently been shown to outperform other metals in plasmon-enhanced thin-film solar cells due to its low damping and strong field localization \cite{Atwater2010}. The present findings confirm this trend in CTS-based heterostructures and suggest that combining Ag with optical cavity engineering and proper buffer layer design (e.g., MoS$_2$) can substantially improve light harvesting.

In summary, this figure reveals that Ag is the optimal top metal for maximizing light absorption in CTS when integrated in a multilayer photonic configuration. This result aligns well with prior work on plasmon-enhanced light trapping in semiconducting thin films and validates the potential of CTS as a cost-effective, non-toxic absorber for next-generation solar and photothermal applications. The performance gap between Ag and Au also reflects the trade-off between optical performance and chemical stability, suggesting that future work may explore protective coatings or alloying to combine the benefits of both.

\section{Conclusions}
We have examined the optimal optical absorption on thin films toward the application in solar cells by using a system of a semi-transparent thin film and a reflective layer embedded on a substrate. It has been shown that for metallic materials such as Au, Ag, Cu, and Al, the maximal absorption could be significantly enhanced thanks to reflection from the second layer. The optimal absorption happens at small thicknesses below 25 nm, and these thicknesses depend on the optical wavelength. An enhancement of more than twice the optical absorption of a single thin film has been obtained, from 20\% to $>$52\% for the case of Au and similar for other materials. This presents an important role for the reflective layer in enhancing the total absorption of the active metallic layer. The system could be fabricated with other modern two-dimensional materials such as nanoparticles (NPs) \cite{LeeOE10,SpinelliJOpt12, WuAPYA17Ag,SrivastavaAPYA18Au, HamedAPYA20Ag} or NPs combined with organic materials \cite{WaketolaAPYA23} to exploit the plasmonic coupling, graphene \cite{JabeenIEEEQm18} to enhance the opto-electrical interaction, or nanogratings \cite{SubhanRSC20} on the thin films themselves, or even recently to increase absorption. Therefore, the results in this study could provide fruitful information for selecting suitable parameters for the system toward optimized absorption on the surface layer, or any layer in the middle as a host for the modified structure, to enhance its energy harvesting functions.

\bibliographystyle{elsarticle-num}
\bibliography{VybibAIP2_a,solar_cell_23}

\end{document}